\documentclass[smallextended]{svjour3}       
\smartqed  
\usepackage{amsmath}
\usepackage{amssymb}
\usepackage{graphicx}          
\usepackage{psfrag}
\usepackage{tikz}
\usepackage{pgfplots}
\usepackage{adjustbox}
\usepackage{xcolor}
\usepackage{mathtools}
\newtheorem{defn}{Definition}

 \journalname{Nonlinear Dynamics}
\begin{document}

\title{Symmetries in the wheeled inverted pendulum mechanism
}


\author{Sneha Gajbhiye, Ravi N. Banavar, Sergio Delgado}


\institute{S. Gajbhiye and R. N. Banavar \at Systems and Control Engineering, Indian Institute of Technology, Bombay \\ \email{sneha@sc.iitb.ac.in; banavar@iitb.ac.in}  
           \and
           S. Delgado \at  Technical University of Munich, Garching, Germany \\
            \email{ s.delgado@tum.de} 
}


\maketitle

\begin{abstract}
The purpose of this article is to illustrate the role of connections and symmetries in the Wheeled Inverted
Pendulum (WIP) mechanism - an underactuated system with rolling constraints - popularized commercially as the Segway, and thereby arrive at a set of simpler dynamical equations that could serve as the starting point for more complex feedback control designs. The first part of the article views the nonholonomic constraints enforced by the rolling assumption as defining an Ehresmann connection on 
a fiber bundle. The resulting equations are the reduced Euler Lagrange equations, which are identical to the Lagrange d'Alembert equations of motion. In the second part we explore conserved quantities, in particular, nonholonomic momenta. To do so, we first introduce the notion of a symmetry group, whose action leaves both the Lagrangian and distribution invariant. We examine two symmetry groups - $SE (2)$ and $SE(2) \times \mathbb{S}^{1}$. The first group leads to the purely kinematic case while the second gives rise to nonholonomic momentum equations. 
\end{abstract}

\section{Introduction}
\label{intro}
\begin{figure}[h]
\centering
\includegraphics[scale=0.5]{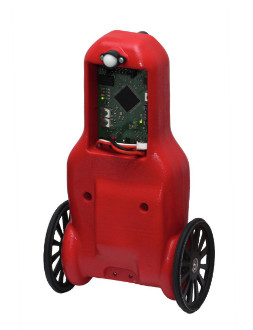}
\caption{$KRT32$- the wheeled inverted pendulum developed at TUM}
\label{prototype}
\end{figure}

The class of nonholonomic systems forms a large and interesting subset of mechanical control systems. Applications include robotics, rolling and locomotive mechanisms. A better understanding of the system's intrinsic structure and properties, at times, simplifies control synthesis. Though the classical approaches, like Lagrange-d'Alembert's principle, yield the equations of motion, geometric approaches exploit underlying properties like symmetry and help understand the structure and intrinsic properties of nonholonomic mechanical systems. \cite{bloch2003} is a comprehensive introduction to these notions. In this article we study the geometric features of one such system, the \textit{Wheeled Inverted Pendulum}, using tools of geometric mechanics. A miniaturized and compact version of the WIP (see Figure (\ref{prototype})) has been designed and developed in the Institute of Automatic Control, TUM. This prototype is currently being used as an experimental test bed for candidate control algorithms.

The Wheeled Inverted Pendulum (WIP) consists of a vertical body with two coaxial driven wheels. Typical applications of the WIP include baggage transportation, commuting and navigation \cite{Segway2014}. The WIP has gained interest in the past several years due to its maneuverability and simple construction (see e.g. \cite{Grasser2002}, \cite{chan2013review}). Other robotic systems based on the WIP are becoming popular as well in the robotic community for human assistance or transportation as can be seen in the works of \cite{Li2012}, \cite{Nasrallah2006}, \cite{Nasrallah2007}, \cite{Baloh2003}, and a commercially available model $Segway$ for human transportation \cite{Segway2014}. The stabilization and tracking control for the WIP is challenging since the system belongs to a class of underactuated mechanical systems (the control inputs are less than the number of configuration variables) and has nonholonomic constraints as well, that arise due to rolling without slipping assumptions on the wheels. Several control laws have been applied to the WIP, mostly using linearized models as can be seen in \cite{blankespoor2004experimental}, \cite{salerno2004control}, \cite{kim2005dynamic}, \cite{Li2012}. In \cite{salerno2003nonlinear}, controllability of the dynamics involving the rotation of the wheels and the pitch of the vertical body (pendulum) were presented and in \cite{salerno2004control} a linear controller was designed for stabilization. In \cite{kim2005dynamic}, the authors presented the exact dynamics of WIP and derived the linear controller. In \cite{pathak2005velocity} and \cite{gans2006visual}, the authors propose a the controller based on partial feedback linearization. \cite{Nasrallah2007} develops a model based on the Euler-Rodrigues parameters and analyzes the controllability of the WIP moving on an inclined plane. However, the geometric structure of the WIP and the consequent aid to feedback design
is yet to be completely exploited. In \cite{delgado2015reduced}, the authors adopt a geometric approach to derive the dynamic model. The aim of this article is to present geometric facets, various group symmetries of the dynamics of this system, in particular the nonholonomic momentum, and finally arrive at a model which would considerably aid in control design.

In mechanical systems with nonholonomic constraints, the configuration space $Q$ is a finite dimensional smooth manifold, the tangent bundle $TQ$ is the velocity phase space, the Lagrangian is a map $L: TQ \longrightarrow \mathbb{R}$ and a smooth distribution $\mathcal{D} \subset TQ$ determines the nonholonomic constraints. Typically, $L$ is the kinetic energy minus the potential energy. So, at a given point of the configuration space, the distribution $\mathcal{D}$ characterizes the allowable velocity directions of the system. The Lagrange-d'Alembert principle, then, yields the equations of motion of the system. The constraints form the horizontal space of the tangent space in a direct sum of two subspaces - termed horizontal and vertical, and this horizontal space is realized through an Ehresmann connection. The dynamics then appear in the reduced Euler-Lagrange form with the constraint forcing terms dependant on the curvature of the connection. More scholarly exposition on this is found in \cite{BKMM}, \cite{bloch2003}, \cite{unicycle}. Often, nonholonomic systems admit a symmetry group, and the action of this group usually makes analysis simpler. The configuration space is then identified ``locally" as the product space of a group and a shape space, with the group being the symmetry group. This modifies the Ehresmann connection to a new connection which is associated with this symmetry. This new connection, which is a principal connection, is termed as a $nonholonomic$ $connection$ and the dynamics is studied on a reduced space or shape space. Hence, using group symmetry one performs Lagrangian reduction and obtains reduced dynamics with a reconstruction equation combined with constraints. The nonholonomic connection, in turn, is realized through two types of connections, one due to the constraints - the $kinematic$ $connection$ - and the second, arising due to the kinetic energy metric, termed as the $mechanical$ $connection$. So, the $nonholonomic$ $connection$ holds information about both the constraints and the dynamics.
The general references on reduction theory with constraints are  \cite{marsden1993reduced}, \cite{marsden2000reduction}, \cite{c2}, \cite{bloch2003}, \cite{BKMM}, \cite{ostrowski_thesis}, \cite{CHMR}, \cite{CMR}. There are three cases to be noticed while computing this principal connection. If the distribution forms the horizontal space, then the principal connection is
realized as a kinematic connection and hence we have fiber (vertical) symmetry which yields the standard momentum conservation in both spatial and body frames \cite{koiller1992reduction}. If the distribution forms the horizontal symmetry, that is, the distribution lies 
in the fiber space, then the equation of motion is in the Euler-Poincar\'{e} form and the momentum is conserved in the spatial frame, for example in the vertical coin \cite{bloch2003}. And the third one is a general case where the distribution 
partially lies both in the horizontal and the vertical spaces. This general case gives rise to a generalized momentum equation wherein the momentum is not necessarily conserved. In \cite{BKMM}, \cite{ostrowski1995mechanics} authors illustrates the Snakeboard example where nonconservation of momentum plays an important role in locomotion. The WIP falls under the category of the general case where the forward motion of the wheels and yaw motion are given by the generalized momentum.

The objective of this paper is to illustrate the symmetries  and conserved quantities inherent in the dynamics of the Wheeled Inverted Pendulum (WIP). The first part of the article uses the Ehresmann connection and formulates the dynamics in the form of
the Lagrange-d'Alembert equation with the Euler-Lagrange equation in the base variables with curvature form. The nonholonomic constraint
distribution forms the horizontal subspace for the Ehresmann connection. In the literature this is called as the kinematic connection. The second formulation uses the notion of a nonholonomic system with symmetry. Here, a symmetry group acts on the WIP configuration space and renders the Lagrangian and distribution invariant. Two types of Lie group symmetries are considered. A connection termed as nonholonomic connection is introduced, which synthesizes the mechanical connection and kinematic connection. This analysis gives rise to momentum equation for group variables and reduced Euler-Lagrange equation for shape variables.  
\section{System description}
The WIP consists of a body of mass $m_b$ (center of mass at a distance $b$ from the wheels rotation axes) mounted on two wheels of radius $r$. Let $m_{W}$ be the mass of the wheels and $d$ be the distance between the wheels. The wheels are directly mounted on the body and are able to rotate independently. Since the wheels are actuated by motors sitting on the body, a tilting motion automatically rotates the wheels through the tilting angle. The body needs to be stabilized in the upper position through a back and forth motion of the system similar to the inverted pendulum on a cart. The set of generalized coordinates describing the WIP are:
\begin{enumerate}
\item Coordinates of the origin of the body-fixed coordinate system in the horizontal plane ($x, y \in \mathbb{R}^2$)
\item Heading angle around the $_Iz$-axis ($\theta \in \mathbb{S}^1$)
\item Tilting angle around the $_Sy$-axis ($\alpha \in \mathbb{S}^1$)
\item Relative rotation angle of each of the wheels with respect to the body around the $_{W_j}y$-axis, which coincides with the $_By$-axis ($\phi_1 \in \mathbb{S}^1$ and $\phi_2 \in \mathbb{S}^1$)
\end{enumerate}
The configuration space ${Q}$ of the system is thus $(\mathbb{R}^2 \times \mathbb{S}^1) \times (\mathbb{S}^1 \times \mathbb{S}^1 \times \mathbb{S}^1) = G \times S$ with $q=(x,y,\theta, \alpha, \phi_{1},\phi_{2} )$. \\
\begin{figure}[ht]
\centering
\includegraphics[width=0.45\textwidth]{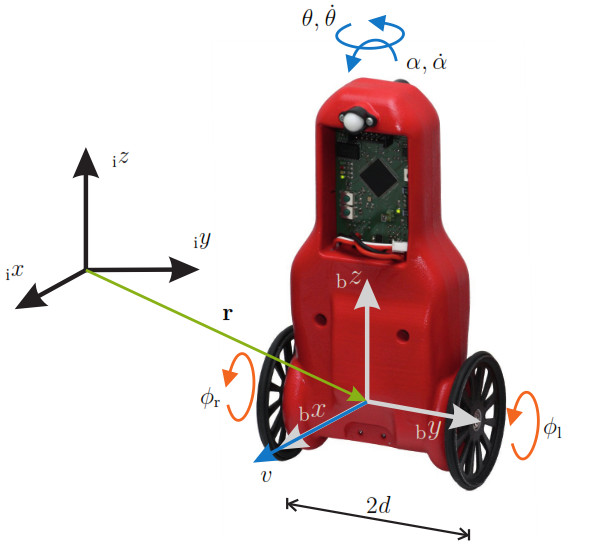}
\caption{The Wheeled Inverted Pendulum}
\label{fig:Skizze2}
\end{figure}
%
Assuming the wheels roll without slipping, the system has nonholonomic constraints given by:
\begin{align}\label{wheel_constraint}
\begin{split}
& \dot{x}_{L} \cos \theta + \dot{y}_{L} \sin \theta = r \dot{\phi}_{1}; \\
 & \dot{x}_{R} \cos \theta + \dot{y}_{R} \sin \theta = r \dot{\phi}_{2}; \\
 & -\dot{x}_{L/R} \sin \theta + \dot{y}_{L/R} \cos \theta =0,
\end{split}
\end{align}
With $x_{L} = x - (d/2) \sin \theta $, $y_{L} = y + (d/2) \cos \theta $, $x_{R} = x + (d/2) \sin \theta $ and $y_{R} = y - (d/2) \cos \theta $ these are equivalent to
\begin{align}
& \dot{x} - r \cos \theta (\dot{\phi}_{1} + \dot{\phi}_{2}) =0, \label{constraint_eqn1} \\
& \dot{y} - r \sin \theta (\dot{\phi}_{1} + \dot{\phi}_{2})=0, \label{constraint_eqn2} \\
& \dot{\theta} - \frac{r}{d} (\dot{\phi}_{2} - \dot{\phi}_{1})=0. \label{constraint_eqn3}
\end{align}
Equations \eqref{constraint_eqn1} and \eqref{constraint_eqn2} are treated as nonintegrable constraints, that is, the translational velocity of the body (chassis) in both $x$ and $y$ directions are completely determined by the angular velocities of the wheels and the body yaw angle. Equation \eqref{constraint_eqn3} is, in fact, a holonomic constraint, which relates the yaw angle with the roll angle of the wheels.. Consider coordinates $q^{i}$
$(i=1,\cdots,n $) for $Q$,  and a set of velocities $\dot{q}$ at point $q$, that defines the tangent space $T_{q}Q$. Let $\mathcal{D}$ be a distribution that describes the kinematic constraints as above. So, at a given point $q$, the distribution $\mathcal{D}$ characterizes the allowable velocity direction of the system, i.e, $\mathcal{D}$ is a collection of linear subspaces $D_{q} \subset T_{q}Q$, $\forall q \in Q$. The nonholonomic constraint can be expressed as $\dot{s}^{a} - A_{\beta}^{a}\dot{r}^{\beta}=0$, where $s = (x,y,\theta)$, $r = (\alpha, \phi_{1}, \phi_{2})$ and 
\begin{equation}
A = \begin{bmatrix}
0 & - r \cos \theta & - r \cos \theta \\
0 & - r \sin \theta & - r \sin \theta \\
0 & \frac{r}{d} & -\frac{r}{d}
\end{bmatrix}.
\end{equation}
The above constraints have a geometric interpretation.
Consider a bundle map $\pi_{Q,S}: Q \longrightarrow S$ which is a submersion and $S$ is termed the
 $base$ $space$, then $T_{q}\pi_{Q,S}$ is a derivative map (onto) at each $q \in Q$ and kernel of $T_{q}\pi_{Q,S}$ at any point forms a vertical space $V_{q}$. This vertical subbundle (of $TQ$) is also referred as the fiber distribution, 
 and is defined as
\begin{equation}
VQ = \cup_{q \in Q} V_{q}Q \;\;\; \;  V_{q}Q = \{ v_{q} \in T_{q}Q | v_{q} \in ker T_{q}\pi \}.
\end{equation}
\begin{defn}
An Ehresmann connection $A$ is defined as a vector-valued one form which splits the tangent space $T_{q}Q$ at every point $q \in Q $  into a vertical and a horizontal space, satisfying: 1) $A_{q}:T_q Q \longrightarrow V_{q}$ and, 2) $A$ is a projection, $A(v_{q}) = v_{q}$ for all $v_{q}\in V_{q}$. For $T_{q}Q=H_{q} \oplus V_{q}$, implying $H_{q}=\mbox{ker}A_{q}$.
\end{defn}
We choose the Ehresmann connection such that the constraint distribution forms the horizontal space. For $X_{q} \in TQ$, we then have the horizontal part of the vector field as $H (X_{q}) = X_{q} - A(q)X_{q}$. In bundle coordinates, $q=(s,r)$, the connection which is a vector-valued one form can be expressed in local coordinates as $A = (ds^{a} + A_{\beta}^{a}(s,r)dr^{\beta})\dfrac{\partial}{\partial s^{a}}$. Therefore, the vertical component is  $(\dot{r},\dot{s}) \mapsto (0, \dot{s} + A(s,r)\dot{r}^{\beta})$ and the horizontal component $(\dot{r},\dot{s}) \mapsto (\dot{r}, - A(s,r)\dot{r})$.

The Lagrangian for the WIP is taken to be the total kinetic energy minus the potential energy and is given by
\begin{align}
L(q,\dot{q} & ) = \frac{1}{2}(m_{b} + 2m_{W})\dot{x}^{2} + \frac{1}{2}(m_{b} + 2m_{W}) \dot{y}^{2} + \frac{1}{2}I_{\theta}(\alpha) \dot{\theta}^{2} \nonumber \\ 
& + \frac{1}{2}(m_{b}b^{2} + I_{Byy}) \dot{\alpha}^{2} + \frac{1}{2} I_{Wyy} (\dot{\phi}_{1}^{2} + \dot{\phi}_{2}^{2}) - m_{b}b \sin \alpha \sin \theta \dot{x}\dot{\theta} \nonumber \\
& + m_{b}b \cos \alpha \cos \theta \dot{\alpha} \dot{x} + m_{b}b \sin \alpha \cos \theta \dot{\theta} \dot{y} + m_{b}b \cos \alpha \sin \theta \dot{\alpha} \dot{y} \nonumber \\
& - m_{b}b g \cos \alpha \label{lagrangian}
\end{align}
where,
\begin{equation*}
I_{\theta}(\alpha) = 2 I_{Wzz} + I_{Bz} \cos^{2} \alpha + 2 m_{W}d^{2} + (I_{Bxx} + m_{B}b^{2}) \sin^{2} \alpha.
\end{equation*}
Given $L : TQ \longrightarrow \mathbb{R}$ and a smooth distribution $\mathcal{D}$ that represents the 
constraints, the Lagrange-d'Alembert principle that yields the equations of motion states that the motion of the system occurs along trajectories that satisfy Hamilton's variational principle where the variations of $L$ are taken along curves which satisfy $\mathcal{D}$, and are assumed to vanish at the endpoints. We now define the constrained Lagrangian by substituting the constraints \eqref{constraint_eqn1}-\eqref{constraint_eqn3} into the Lagrangian as
\begin{equation}
L_{c}(\alpha, \dot{\alpha}, \dot{\phi}_{1}, \dot{\phi}_{2}) = L (\theta,\alpha, r \cos \theta (\dot{\phi}_{1} + \dot{\phi}_{2}), r\sin \theta (\dot{\phi}_{1} + \dot{\phi}_{2}), \frac{r}{d}(\dot{\phi}_{2} - \dot{\phi}_{2}))  \nonumber   \\ 
\end{equation}
yielding $L_{c}$ as
\begin{align*}
L_{c} & = \frac{1}{2} a_{1} \dot{\phi}_{1}^{2} + \frac{1}{2}a_{1} \dot{\phi}_{2}^{2} +  \frac{1}{2} c~ \dot{\alpha}^{2} + \frac{1}{2}a_{2} \dot{\alpha} (\dot{\phi}_{1} + \dot{\phi}_{2}) +  a_{3} \dot{\phi}_{1}\dot{\phi}_{2} - m_{b}bg\cos \alpha
\end{align*}
where
\begin{align*}
& a_{1} = \left( \frac{1}{4}(m_{b} + 2m_{W})r^{2} + \frac{r^{2}}{4d^{2}} I_\theta(\alpha) + I_{Wyy} \right);\\
& a_{2} = \left( (m_{b} + 2m_{W})r^{2} + m_{b}b \cos \alpha \right); \\
& a_{3} = \left( \frac{1}{4}(m_{b} + 2m_{W})r^{2} - \frac{r^{2}}{4d^{2}} I_\theta(\alpha) \right); \\
& c = (m_{b}b^{2} + I_{Byy}).
\end{align*}
Following \cite{BKMM}, the equations of motion in terms of the constrained Lagrangian $L_c$, termed as reduced Euler-Lagrange equations, are given by
%
\begin{align}
\frac{d}{dt}\left( \frac{\partial L_{c}}{\partial \dot{\alpha}} \right) - \frac{\partial L_{c}}{\partial \alpha} + A_{\alpha}^{a}\frac{\partial L_{c}}{\partial s^{a}} = - \frac{\partial L}{\partial \dot{s}^{b}} \left( B_{\alpha \alpha}^{b}\dot{\alpha} + B_{\alpha \phi_{1}}^{b}\dot{\phi_{1}} + B_{\alpha \phi_{2}}^{b}\dot{\phi_{2}} \right), \label{eq:2:1} \\
\frac{d}{dt}\left( \frac{\partial L_{c}}{\partial \dot{\phi_{1}}} \right) + A_{\phi_{1}}^{a}\frac{\partial L_{c}}{\partial s^{a}} = - \frac{\partial L}{\partial \dot{s}^{b}} \left( B_{\phi_{1} \alpha}^{b}\dot{\alpha} + B_{\phi_{1} \phi_{1}}^{b}\dot{\phi_{1}} + B_{\phi_{1} \phi_{2}}^{b}\dot{\phi_{2}}\right), \label{eq:2:2} \\
\frac{d}{dt}\left( \frac{\partial L_{c}}{\partial \dot{\phi}_{2}} \right) + A_{\phi_{2}}^{a}\frac{\partial L_{c}}{\partial s^{a}} = - \frac{\partial L}{\partial \dot{s}^{b}} \left( B_{\phi_{2} \alpha}^{b}\dot{\alpha} + B_{\phi_{2} \phi_{1}}^{b}\dot{\phi_{1}} + B_{\phi_{2} \phi_{2}}^{b}\dot{\phi_{2}} \right) \label{eq:2:3}
\end{align}
where the curvature $B_{\beta \gamma}^{b}$ is
\begin{equation}
B_{\beta \gamma}^{b} = \left( \frac{\partial A_{\beta}^{b}}{\partial r^{\gamma}} - \frac{\partial A_{\gamma}^{b}}{\partial r^{\beta}} + A_{\beta}^{a} \frac{\partial A_{\gamma}^{b}}{\partial s^{a}} - A_{\gamma}^{a}\frac{\partial A_{\beta}^{b}}{\partial s^{a}} \right) 
\end{equation}
where $A_{\alpha}^{a}$ are the coordinate expression of the Ehresmann connection on the tangent bundle defined by the constraints.
%
The Ehresmann connection in coordinates is
\begin{eqnarray}\label{ehresmann_connection}
A_{\alpha}^{x} =0, & A_{\phi_{1}}^{x}= - r \cos \theta, & A_{\phi_{2}}^{x}= -r \cos \theta, \nonumber \\
A_{\alpha}^{y} =0, & A_{\phi_{1}}^{y}= - r \sin \theta, & A_{\phi_{2}}^{y}= -r \sin \theta, \nonumber \\
A_{\alpha}^{\theta} =0, & A_{\phi_{1}}^{\theta}= \frac{r}{d}, & A_{\phi_{2}}^{\theta}= - \frac{r}{d}. 
\end{eqnarray}
and the coefficients $B_{\beta \gamma}^{b}$ are given by
\begin{align*}
& B_{\alpha \alpha}^{b} = B_{\alpha \phi_{1} }^{b} = B_{\alpha \phi_{2}}^{b} = B_{\phi_{1} \alpha}^{b} = B_{\phi_{2} \alpha}^{b} = B_{\phi_{1} \phi_{1}}^{b} = B_{\phi_{2} \phi_{2}}^{b} = 0, \\
& B_{\phi_{1} \phi_{2}}^{b} = A_{\phi_{1}}^{\theta} \frac{\partial A_{\phi_{2}}^{b}}{\partial \theta} - A_{\phi_{2}}^{\theta} \frac{\partial A_{\phi_{1}}^{b}}{\partial \theta}, \quad  B_{\phi_{2} \phi_{1}}^{b} = A_{\phi_{2}}^{\theta} \frac{\partial A_{\phi_{1}}^{b}}{\partial \theta} - A_{\phi_{1}}^{\theta} \frac{\partial A_{\phi_{2}}^{b}}{\partial \theta}.
\end{align*}
The equations of motion, calculated from \eqref{eq:2:1}-\eqref{eq:2:3} are
\begin{align}\label{euler_lagrange}
\begin{split}
& \frac{d}{dt}\left( \frac{\partial L_{c}}{\partial \dot{\alpha}} \right) - \frac{\partial L_{c}}{\partial \alpha} = 0, \\
& \frac{d}{dt} \left( \frac{\partial L_{c}}{\partial \dot{\phi}_{1}} \right) = - 2 m_{b}b \frac{r^{2}}{d} \sin \alpha \left( \frac{r}{d} (\dot{\phi}_{2} - \dot{\phi}_{1}) \right)  \dot{\phi}_{2} + \tau_{1} \\
& \frac{d}{dt} \left( \frac{\partial L_{c}}{\partial \dot{\phi}_{2}} \right) = 2 m_{b}b \frac{r^{2}}{d} \sin \alpha \left( \frac{r}{d} (\dot{\phi}_{2} - \dot{\phi}_{1}) \right)  \dot{\phi}_{1} + \tau_{2} 
\end{split}
\end{align}
where $\tau_{1}$ and $ \tau_{2}$ are the  respective torques on the two wheels. These equations of motion are identical to the one given in \cite{delgado2015reduced}. In section 3, we show that using $\mathbb{SE}(2)$ symmetry yields a principal kinematic case and, the principal connection in such a case is equivalent to the Ehresmann connection defined above. 
For a completeness of exposition, the preliminary notions of a nonholonomic system with symmetry as developed in \cite{bloch2003}, \cite{BKMM} are presented in the Appendix.  We now illustrate these tools on the WIP system.

\section{Symmetries in the WIP mechanism}
We make two choices for the group action: 
\begin{itemize}
\item $G_{1}= \mathbb{SE}(2)$ (the position on the plane and the yaw angle) and 
\item $G_{2}= \mathbb{SE}(2) \times \mathbb{S}^{1}$ (the position on the plane and the yaw angle, and the roll of the wheels) .
\end{itemize}
In the first case we illustrate that the principal connection is equivalent to the Ehersmann connection and the equations of motion are given by the reduced Euler-Lagrange equations. In the second case we first identify the configuration variable as $q=(x,y,\theta, \alpha, \phi = \phi_{1} + \phi_{2})$ and then choose the group action of $ \mathbb{SE}(2) \times \mathbb{S}^{1}$ on this $Q$. 
This choice of $Q$ comes from the fact that instead of the absolute wheel angles $\phi_{1}$ and $\phi_{2}$ we take the difference $\alpha = \phi_{1} - \phi_{2}$ which represent the yaw angle,  and the sum $\phi = \left( \frac{\phi_{1} + \phi_{2}}{2} \right)$ which permits us to calculate the forward distance traversed as $r \phi$. This modified choice also comes from the ultimate control synthesis objective where one wants to control the forward and yaw velocity of the WIP.
In this case we illustrate the nonholonomic momentum and derive the  reduced nonholonomic Lagrange-d'Alembert equations. \\
\textbf{Case I:} Consider the Lie group $G_{1}= \mathbb{SE}(2)$ and the symmetry in $s=(x,y,\theta)$ variables of the system. The action by the group element $(\bar{x},\bar{y},\bar{\theta})$ is given by
$$(x,y,\theta, \alpha, \phi_{1}, \phi_{2}) \longmapsto (x \cos \bar{\theta} - y \sin \bar{\theta} + \bar{x}, x \sin \bar{\theta} + y \cos \bar{\theta} + \bar{y}, \theta + \bar{\theta}, \alpha, \phi_{1}, \phi_{2}) $$
The tangent space to the $\mathbb{SE}(2)$ group orbit is given by
\begin{equation}
T_{q}\mbox{Orb}(q) = \mbox{span} \lbrace \frac{\partial}{\partial x}, \frac{\partial}{\partial y}, \frac{\partial}{\partial \theta} \rbrace
\end{equation}
The Lagrangian \eqref{lagrangian} and constraints \eqref{wheel_constraint} are invariant under the action of $G_{1}$. The vector fields $X_{1}, X_{2}, X_{3}$ that are the local generators for the constrained distribution $\mathcal{D}$ and are given by
\begin{align*}
& X_{1} = \cos \theta \frac{\partial}{\partial x} - \sin \theta \frac{\partial}{\partial y} + \frac{1}{r} \frac{\partial}{\partial \phi_{1}} + \frac{1}{r} \frac{\partial}{\partial \phi_{2}}, \\
& X_{2} = \frac{\partial}{\partial \alpha}, \\
& X_{3} = \frac{\partial}{\partial \theta} + \frac{d}{r} \frac{\partial}{\partial \phi_{1}} - \frac{d}{r} \frac{\partial}{\partial \phi_{2}} 
\end{align*}
therefore, 
\begin{equation}
\mathcal{D}_{q} = \mbox{span} \lbrace X_{1}, X_{2}, X_{3} \rbrace.
\end{equation}
The intersection of the tangent space to the orbit with the constrained distribution $\mathcal{S}_{q}=\mathcal{D}_{q} \cap T_{q}\mbox{Orb}(q) = \lbrace 0 \rbrace $ and the components of curvature are independent of $x$ and $y$. This is the 
{\it principal kinematic case}, in which there is a principal connection whose horizontal space is spanned by the distribution $\mathcal{D}$. The projection on the vertical space defines the Ehresmann connection and since the distribution is invariant under the group action, the principal connection related to the Ehresmann connection as $A = \mathcal{A}_{Q}$, given in \eqref{ehresmann_connection}. The system is reduced from $TQ$ to $TQ/G_{1}=\mathfrak{se}(2) \times T \mathbb{S}^{1}\times T \mathbb{S}^{1} \times T \mathbb{S}^{1} $ by the group action. Substituting the constraint $\dot{s} = - A(s,r)\dot{r}$ from \eqref{constraint_eqn1}-\eqref{constraint_eqn3}, the reduced equations of motion are obtained on $ \mathcal{D}/G_{1}= T \mathbb{S}^{1}\times T \mathbb{S}^{1} \times T \mathbb{S}^{1}$. The reduced constrained Lagrangian is
\begin{align}\nonumber
\begin{split}
l_{c} (\alpha,\dot{\alpha},\dot{\phi}_{1},\dot{\phi}_{2})
 & =\mathit{l}(r\cos \theta (\dot{\phi}_{1} + \dot{\phi}_{2}),r\sin \theta ( \dot{\phi}_{1} + \dot{\phi}_{2}), \frac{r}{d}(\dot{\phi}_{2} - \dot{\phi}_{2}))
\end{split}
\end{align}
There is no momentum equation and correspondingly no body velocity. The equation of motion is the reduced Euler-Lagrange equation given in \eqref{euler_lagrange} with the reconstruction equation being $\dot{s} = - A(s,r)\dot{r}$.\\
\textbf{Case II:} Consider the group action $G_{2} = \mathbb{SE}(2) \times \mathbb{S}^{1}$ on $Q = \mathbb{SE}(2) \times \mathbb{S}^{1} \times \mathbb{S}^{1}$. The
group here denotes the $(x,y)$ position, heading angle $\theta$ and $\phi$ sum of wheel angles. Let $\dot{\phi} =  \left( \frac{\dot{\phi}_{1} + \dot{\phi}_{2}}{2} \right)$ be sum of wheel velocity results in the forward velocity $r \dot{\phi}$ of the cart. With this let the configuration space is now identify as $Q = \mathbb{SE}(2) \times \mathbb{S}^{1} \times \mathbb{S}^{1}$ with configuration variables as $q=(x,y,\theta, \alpha,\phi)$. Then the left action of $G_{2}$ on $Q$ is given as
%
%
$$\Phi:(x,y,\theta, \alpha, \phi) \longmapsto (x \cos \bar{\theta} - y \sin \bar{\theta} + \bar{x}, x \sin \bar{\theta} + y \cos \bar{\theta} + \bar{y}, \theta + \bar{\theta}, \alpha, \phi + \bar{\phi}) $$
where $(\bar{x},\bar{y},\bar{\theta}) \in \mathbb{SE}(2)$ and $\bar{\phi} \in \mathbb{S}^{1}$. The left action of $G$ on the tangent-lifted coordinates of the manifold $Q$ is
\begin{align*}
T\Phi_{(\bar{x},\bar{y},\bar{\theta},\bar{\phi})} & :(x,y,\theta, \alpha, \phi, \dot{x},\dot{y},\dot{\theta},\dot{\alpha},\dot{\phi}) \longrightarrow \nonumber \\
& (x \cos \bar{\theta} - y \sin \bar{\theta} + \bar{x}, x \sin \bar{\theta} + y \cos \bar{\theta} + \bar{y}, \theta + \bar{\theta}, \alpha,  \nonumber \\
& \phi + \bar{\phi}, \dot{x}\cos \bar{\theta} - \dot{y}\sin \bar{\theta}, \dot{x}\sin \bar{\theta} + \dot{y}\cos \theta, \dot{\theta},\dot{\alpha},\dot{\phi}) \nonumber
\end{align*}
%
%
The Lagrangian $L$ of the system is,
\begin{align}
L(q,\dot{q}) & = \frac{1}{2}(m_{b} + 2m_{W})(\dot{x}^{2} + \dot{y}^{2}) + \frac{1}{2}\left( I_{\theta}(\alpha) + \frac{d^{2}}{2r^2}I_{Wyy} \right) \dot{\theta}^{2} \nonumber \\
& + \frac{1}{2}(m_{b}b^{2} + I_{Byy}) \dot{\alpha}^{2} + \frac{1}{2} I_{Wyy}2 \dot{\phi}^{2} + m_{b}b \sin \alpha \dot{\theta} (-\sin \theta \dot{x} \nonumber \\
& + \cos \theta \dot{y} ) + m_{b}b \cos \alpha \dot{\alpha} ( \cos \theta \dot{x} + \sin \theta \dot{y}) - m_{b}b g \cos \alpha \label{lagrangian2}
\end{align}
and the nonholonomic constraint is
\begin{equation}\label{new_constraint}
\dot{x} - r \cos \theta \dot{\phi} =0, \quad \quad \dot{y} - r \sin \theta \dot{\phi} =0.
\end{equation}
It is easily proved that the Lagrangian $L$ and distribution $\mathcal{D}$ are invariant under the action of the group $G$.  \\
Substituting the constraints in the Lagrangian \eqref{lagrangian2}, the constrained Lagrangian $L_{c}$ is determined as 
\begin{align}
L_{c} & = (m_{b} + 2m_{W})r^{2} \dot{\phi}^{2} + rm_{b}b \cos \alpha \dot{\phi} \dot{\alpha} + \frac{1}{2} \left( I_{\theta} + \frac{d^{2}}{2r^{2}} I_{Wyy} \right) \dot{\theta}^{2} \nonumber \\
& + \frac{1}{2} (m_{b}b^{2} + I_{B} ) \dot{\alpha} + \frac{1}{2} 2I_{Wyy} \dot{\phi}^{2}
\end{align}
The tangent space to the $G_2$ orbit is
\begin{equation}
T_{q}\mbox{Orb}(q) = \mbox{span}\lbrace \frac{\partial}{\partial x}, \frac{\partial}{\partial y}, \frac{\partial}{\partial \theta }, \frac{\partial}{\partial \phi} \rbrace
\end{equation}
and the constraint distribution is given by $\mathcal{D}_{q} = \mbox{span}\{ X_{1},X_{2},X_{3}\}$.
%
with
\begin{align*}
& X_{1} = \cos \theta \frac{\partial}{\partial x} - \sin \theta \frac{\partial}{\partial y} + \frac{1}{r} \frac{\partial}{\partial \phi};  \quad X_{2} = \frac{\partial}{\partial \alpha}; \quad X_{3} = \frac{\partial}{\partial \theta}.
\end{align*}
The constraint fiber distribution $S_{q}$ is calculated as
\begin{equation}
\mathcal{S}_{q} = \mathcal{D}_{q} \cap T_{q}\mbox{Orb}(q) = \lbrace \cos \theta \frac{\partial}{\partial x} + \sin \theta \frac{\partial}{\partial y} + \frac{1}{r} \frac{\partial}{\partial \phi}, \quad \frac{\partial}{\partial \theta } \rbrace
\end{equation}
For obtaining the corresponding momentum equation, we consider the bundle whose fibers span the tangent vectors in $\mathcal{S}_{q}$ and choose a section of this bundle. Consider $\mathfrak{g}=\mathfrak{se}(2) \times \mathbb{R}$ the Lie algebra of $G_{2}$. The generators corresponding to the Lie algebra elements can be represented in standard basis in $\mathbb{R}^{4}$ as
$$(1,0,0,0)_{Q}=\frac{\partial}{\partial x}, \quad (0,1,0,0)_{Q}=\frac{\partial}{\partial y},$$
$$ (0,0,1,0)_{Q}=-y \frac{\partial}{\partial x} + x \frac{\partial}{\partial y} + \frac{\partial}{\partial \theta}, \quad (0,0,0,1)_{Q}= \frac{\partial}{\partial \phi}$$
where the first two components represent translations, the  third is the yawing motion and the fourth being the rolling motion. Therefore, to obtain the section of $\mathcal{S}_{q}$ given by vector fields
\begin{align}
& (\xi^{q}_{1})_{Q} = r \cos \theta \frac{\partial}{\partial x} + r \sin \theta \frac{\partial}{\partial y} +  \frac{\partial}{\partial \phi}, \quad \mbox{and} \quad  (\xi^{q}_{2})_{Q} = \frac{\partial}{\partial \theta}
\end{align}
and 
the corresponding Lie algebra elements are
\begin{equation}
\xi^{q}_{1} = \left( r \cos \theta, r \sin \theta,0,1 \right)  \mbox{ and } \xi^{q}_{2} = ( y, -x,1, 0 ).
\end{equation}
We have two the nonholonomic momenta corresponding to the two infinitesimal generators in $\mathcal{S}_{q}$. The nonholonomic momentum in the body representation are calculated from \eqref{momentum_formula1} and \eqref{momentum_formula} as
\begin{equation}
p_{i}(\xi^{q}) = \frac{\partial L}{\partial \dot{q}^{i}} (\xi^{q}_{i})_{Q}
\end{equation}
%
%
which yields
\begin{align}
p_{1}& (\xi^{q}_{1})  = \frac{\partial L}{\partial \dot{q}} (\xi^{q}_{1})_{Q} \nonumber \\
&= \langle \left( (m_{b}+2m_{W}) \dot{x} - m_{b}b \sin \alpha \sin \theta \dot{\theta} + m_{b}b \cos \alpha \cos \theta \dot{\alpha}, \right. \nonumber \\
& \left. (m_{b}+2m_{W}) \dot{y} + m_{b}b \sin \alpha \cos \theta \dot{\theta} + m_{b}b \cos \alpha \sin \theta \dot{\alpha}, \right. \nonumber \\
& \left. (I_{\theta} + \frac{d^{2}}{2r^{2}} I_{Wyy})\dot{\theta} + m_{b}b \sin \alpha (- \sin \theta \dot{x} + \cos \theta \dot{y}), (m_{b}b^{2} + I_{Byy}) \dot{\alpha} \right. \nonumber \\
& \left. + m_{b}b \cos \alpha (\cos \theta \dot{x} + \sin  \theta \dot{y}), 2 I_{Wyy} \dot{\phi} \right); (r \cos \theta, r \sin \theta,0,0,1) \rangle \nonumber \\
& = \left[ (m_{b} + 2m_{W})r^{2} + 2I_{Wyy} \right] \dot{\phi} + rm_{b}b \cos \alpha \dot{\alpha} \label{momentum_1}
\end{align}
and 
\begin{align}
p_{2}(\xi^{q}_{2}) & = \frac{\partial L}{\partial \dot{q}} (\xi^{q}_{2})_{Q} = \langle \left( (m_{b}+2m_{W}) \dot{x} - m_{b}b \sin \alpha \sin \theta \dot{\theta} \right. \nonumber \\
& \left. + m_{b}b \cos \alpha \cos \theta \dot{\alpha}, (m_{b}+2m_{W}) \dot{y} + m_{b}b \sin \alpha \cos \theta \dot{\theta} \right. \nonumber \\
& \left. + m_{b}b \cos \alpha \sin \theta \dot{\alpha},  (I_{\theta} + \frac{d^{2}}{2r^{2}} I_{Wyy})\dot{\theta} \right. \nonumber \\
& \left. + m_{b}b \sin \alpha (- \sin \theta \dot{x} + \cos \theta \dot{y}), (m_{b}b^{2} + I_{Byy}) \dot{\alpha} \right. \nonumber \\
& \left. + m_{b}b \cos \alpha (\cos \theta \dot{x} + \sin  \theta \dot{y}), 2 I_{Wyy} \dot{\phi} \right); (0, 0,1,0,0) \rangle \nonumber \\
& = \left[  I_{\theta}(\alpha)  + \frac{d^{2}}{2r^{2}}I_{W_{yy}} \right] \dot{\theta} \label{momentum_2}
\end{align}
From the nonholonomic momenta calculated in \eqref{momentum_1} and \eqref{momentum_2}, the nonholonomic momentum equations are evaluated as
\begin{align}
\frac{d}{dt} p_{1}(\xi^{q}_{1}) & = \frac{\partial L}{\partial \dot{q}} \left[ \frac{d}{dt}(\xi^{q}_{1}) \right]_{Q} = - \left( (m_{b} + 2m_{W})\dot{x} - m_{b}b \sin \alpha \dot{\theta}\sin \theta \right. \nonumber \\
& \left. + m_{b}b \cos \alpha \dot{\alpha} \cos \theta \right)r \sin \theta \dot{\theta} + \left( (m_{b} + 2m_{W}) \dot{y} \right. \nonumber \\
& \left. +  m_{b}b \sin \alpha \dot{\theta} \cos \theta + m_{b}b \cos \alpha \dot{\alpha} \sin \theta \right)r \cos \theta \dot{\theta} \nonumber \\
& = m_{b} r b \sin \alpha \dot{\theta}^{2},
\end{align}
\begin{align}
\frac{d}{dt} p_{2}(\xi^{q}_{2}) & = \frac{\partial L}{\partial \dot{q}} \left[ \frac{d}{dt}(\xi^{q}_{2}) \right]_{Q} = \left((m_{b} + 2m_{W})\dot{x} - m_{b}b \sin \alpha \dot{\theta}\sin \theta \right. \nonumber \\ 
& \left. + m_{b}b \cos \alpha \dot{\alpha} \cos \theta \right) \dot{y} - \left( (m_{b} + 2m_{W}) \dot{y} +  m_{b}b \sin \alpha \dot{\theta} \cos \theta \right. \nonumber \\
& \left. + m_{b}b \cos \alpha \dot{\alpha} \sin \theta \right) \dot{x} \nonumber \\
& = - m_{b} b r \sin \alpha \dot{\phi} \dot{\theta}.
\end{align}
Eliminating $\dot{\phi}$ and $\dot{\theta}$ using equations (\ref{momentum_1}) and (\ref{momentum_2}), the momentum dynamics are expressed as
\begin{align}
& \dot{p}_{1} = \frac{m_{b} r b \sin \alpha}{ [f(\alpha)]^{2}} p_{2}^{2} , \label{momentum_dyn1} \\
& \dot{p}_{2} = - \frac{m_{b} b r \sin \alpha p_{2}}{ f(\alpha) h } \left[ m_{b} b \cos \alpha \dot{\alpha} + p_{1} \right], \label{momentum_dyn2}
\end{align}
with
\begin{align*}
h =  \left( (m_{b} + 2 m_{w})r^{2} + 2I_{W_{yy}} \right) \mbox{  and  }
f(\alpha) = \left(  I_{\theta}(\alpha)  + \frac{d^{2}}{2r^{2}} I_{W_{yy}} \right).
\end{align*}
and 
\begin{equation}
I_{\theta}(\alpha) = 2 I_{Wzz} + I_{Bz} \cos^{2} \alpha + 2 m_{W}d^{2} + (I_{Bxx} + m_{B}b^{2}) \sin^{2} \alpha.
\end{equation}
This completes the momentum equations computation for the group variables corresponding to $\mathcal{S}_{q}$. Now we calculate the  the dynamic equation governing the shape variable $\alpha$ given in \eqref{shape_dynamics}.
\subsection{Shape dynamics of the WIP under $G_{2}$ action}
To explicitly express the shape dynamics, the reduced Lagrangian and the constrained reduced Lagrangian are computed as follows.
\subsubsection{Reduced Lagrangian and constrained reduced Lagrangian}
The rolling constraint \eqref{new_constraint} is now expressed in the body coordinate frame as
\begin{equation}
\xi_{1} = r\xi_{4}; \quad \quad \quad \xi_{2} = 0.
\end{equation}
where $\xi = g^{-1}\dot{g} \in \mathfrak{se}(2) \times \mathbb{R}$ is the (left-invariant) body angular velocity, expressed by
\begin{equation}
\begin{bmatrix}
\xi_{1} \\
\xi_{2} \\
\xi_{3} \\
\xi_{4}
\end{bmatrix} = 
\begin{bmatrix}
 \cos \theta \dot{x} + \sin \theta \dot{y} \\
 - \sin \theta \dot{x} + \cos \theta \dot{y} \\
\dot{\theta} \\
\dot{\phi}
\end{bmatrix}.
\end{equation}
\begin{theorem}\label{thm2}
With $(L, \mathcal{D})$ and $G_{2}$ as the group action, the  constrained reduced Lagrangian is
\begin{align}
\begin{split}
l_{c}(\alpha, & \dot{\alpha},\xi) = \frac{1}{2} \left((m_{b} + 2m_{w})r^{2} + 2 I_{Wyy} \right) \xi^{2}_{4} + rm_{b}b \cos \alpha \dot{\alpha} \xi_{4} \\
& + \frac{1}{2}(I_{\theta} + \frac{d^{2}}{2r^{2}} I_{Wyy}) \xi^{2}_{3} + \frac{1}{2}(m_{b}b^{2} + I_{B}) \dot{\alpha}^{2} - m_{b}gb \cos \alpha.
\end{split}
\end{align}
\end{theorem}
\textit{Proof}: When the Lagrangian $L$ and the distribution $\mathcal{D}$ are invariant under the action of a group $G$, the system is reduced to the quotient space $\mathcal{D}/G$. From a system($L$, $\mathcal{D}$) on $TQ$, a reduced Lagrangian $\mathit{l}: \mathfrak{g} \times T \mathbb{S}^{1} \longrightarrow \mathbb{R} $ is calculated as
\begin{align*}
l(\alpha,\dot{\alpha},\xi) & = \frac{1}{2} (m_{b} + 2m_{w}) \xi^{2}_{1} - m_{b}b \sin \alpha \dot{\theta} \xi_{2} + m_{b}b \cos \alpha \dot{\alpha} \xi_{1} \\
& + \frac{1}{2} (m_{b} + 2m_{W}) \xi^{2}_{2} + \frac{1}{2}(I_{\theta} + \frac{d^{2}}{2r^{2}} I_{Wyy}) \xi^{2}_{3} \\
& + \frac{1}{2}(m_{b}b^{2} + I_{B}) \dot{\alpha}^{2} + \frac{1}{2} (2 I_{Wyy}) \xi^{2}_{4} - m_{b}gb \cos \alpha
\end{align*}
where $\xi = g^{-1}\dot{g} \in \mathfrak{se}(2) \times \mathbb{R} $. The constraint reads $\xi_{1} = r\xi_{4} $ and $\xi_{2} =0$ (this eliminates two variables of the Lie algebra.) The constrained reduced Lagrangian $l_{c}: \mathfrak{so}(2) \times \mathbb{R} \times T \mathbb{S}^{1} \longrightarrow \mathbb{R}$ is
\begin{align}\nonumber
\begin{split}
l_{c}(\alpha,\dot{\alpha},\xi) &= \frac{1}{2} \left((m_{b} + 2m_{w})r^{2} + 2 I_{Wyy} \right) \xi^{2}_{4} + r m_{b}b \cos \alpha \dot{\alpha} \xi_{4}  \\
& + \frac{1}{2}(I_{\theta} + \frac{d^{2}}{2r^{2}} I_{Wyy}) \xi^{2}_{3} + \frac{1}{2}(m_{b}b^{2} + I_{B}) \dot{\alpha}^{2} - m_{b}gb \cos \alpha 
\end{split}
\end{align}
$\hfill \square $\\
Notice that, the nonholonomic momentum \eqref{momentum_1}-\eqref{momentum_2} can also be obtained from the 
constrained reduced Lagrangian as
\begin{equation}
p_{1} = \frac{\partial l_{c}}{\partial \dot{\phi}} \quad \quad \mbox{and} \quad p_{2} = \frac{\partial l_{c}}{\partial \dot{\theta}}.
\end{equation}

\begin{claim}\label{claim1}
The nonhlonomic connection regarded as a principal connection $A^{nhc}$ and is used to define the constraint equations of the form:
\begin{equation}
A^{nhc}(\dot{q}) = g^{-1}\dot{g} + \mathcal{A}(\alpha)\dot{\alpha} = \Gamma(\alpha)p
\end{equation}
where $\mathcal{A} ( \cdot)$ denotes the local form of the nonholonomic connection given as,
\begin{equation}
\mathcal{A}(\alpha) = r \left( \frac{m_{b}b}{h} \cos \alpha \right) \mbox{d} \alpha e_{1} + \left( \frac{m_{b}b}{h} \cos \alpha \right) \mbox{d}\alpha e_{4}, 
\end{equation}
and 
\begin{equation}
\Gamma(\alpha) =  \frac{r}{h} e_{1} + \frac{1}{f(\alpha)} e_{3} + \frac{1}{h}e_{4}.
\end{equation}
\end{claim}
\textit{Proof:} The kinematic constraints and the momentum equation are given by
\begin{align}
& \left.
\begin{array}{l l l}
& \dot{x} - r \cos \theta \dot{\phi}  = 0, \\
& \dot{y} - r \sin \theta \dot{\phi}  = 0,  
\end{array} \right \}  
& \mbox{Nonholonomic constraints} \nonumber \\
& \left.
\begin{array}{l l}
& p_{1} = h \dot{\phi} + m_{b}b r \cos \alpha \dot{\alpha}, \\
& p_{2} = f(\alpha) \dot{\theta}. 
\end{array} \right \}  
& \mbox{Momentum} \nonumber
\end{align}
Adding and subtracting first two equation then substituting for $\dot{\phi}$ from third equation, we can write 
\begin{equation}
\begin{bmatrix}
 (\cos \theta \dot{x} + \sin \theta \dot{y}) \\
 (- \sin \theta \dot{x} + \cos \theta \dot{y}) \\
\dot{\theta} \\
\dot{\phi}
\end{bmatrix} + 
\begin{bmatrix}
\frac{r m_{b}b }{h} \cos \alpha \dot{\alpha} \\
0\\
0\\
\frac{m_{b}b }{h} \cos \alpha \dot{\alpha}
\end{bmatrix}  = 
\begin{bmatrix}
 \frac{r}{h}p_{1} \\
0\\
\frac{1}{f(\alpha)} p_{2} \\
\frac{1}{h} p_{1}
\end{bmatrix}
\end{equation}
These equations have the form
\begin{equation}\label{nonholo_connection}
g^{-1}\dot{g} + \mathcal{A}(\alpha)\dot{\alpha} = \Gamma(\alpha)p,
\end{equation}
where,
\begin{align*}
& \mathcal{A}(\alpha) = \frac{r m_{b}b }{h} \cos \alpha e_{1} \, \mbox{d}\alpha + \frac{m_{b}b }{h} \cos \alpha e_{4} \, \mbox{d} \alpha \\
& \Gamma(\alpha) =  \frac{r}{h} e_{1} + \frac{1}{f(\alpha)} e_{3} + \frac{1}{h}e_{4}.
\end{align*}
Furthermore, if the momenta are given by $p_{1}$ and $p_{2}$, then the allowable trajectories for the system must satisfy
\begin{equation}
A^{nhc}(q) \cdot \dot{q}= \Gamma(\alpha)p= \begin{bmatrix}
\frac{r}{h} & 0 \\
0 & 0 \\
0 & \frac{1}{f(\alpha)} \\
\frac{1}{h} & 0
\end{bmatrix}
\begin{bmatrix}
p_{1} \\
p_{2}
\end{bmatrix}.
\end{equation}
The above equation specifies the trajectories along the fiber in terms of pitch angle (base variable) and momentum variables.
$ \hfill \square$ \\
With this, the dynamics of the WIP system is now calculated.
\begin{theorem}\label{thm1}
Given the invariance of the Lagrangian and the constraints, the flow of $(p_{1},p_{2})$ is independent of the group variable and is governed by the generalized momentum equation,
\begin{align}
& \dot{p}_{1} = p_{2} \rho_{pp}(\alpha) p_{2}, \\
& \dot{p}_{1} = \sigma_{\dot{\alpha}p}(\alpha) \dot{\alpha}p_{2} + \sigma_{pp}p_{1}p_{2}.
\end{align}
where $\rho_{pp}= (m_{b}rb \sin \alpha)/f^{2}(\alpha)$, $\sigma_{\dot{\alpha}p} = (m_{b}^{2}b^{2}r \sin 2 \alpha )/2f(\alpha)h$ and $(m_{b}br \sin \alpha )/2f(\alpha)h$. Further, using the constrained reduced Lagrangian, the shape dynamics is written as
\begin{align}\label{shape_dynamics_eqn}
& \left( m_{b}b^{2} + I_{B} - \frac{m_{b}^{2}b^{2}\cos^{2} \alpha}{h} \right) \ddot{\alpha} = - \frac{m_{b} b \sin 2 \alpha}{2h} \dot{\alpha}^{2} \nonumber \\
& \quad + \frac{1}{2} \frac{\left(m_{b}^{2}b^{2} r\sin 2 \alpha - h f^{\prime}(\alpha) \right)}{h f(\alpha)^{2}}p_{2}^{2}  - m_{b}gb \sin \alpha
\end{align}
\end{theorem}
\textit{Proof:} The proof of this follows by taking $\Omega = \Gamma(\alpha)p$, then from equation \eqref{nonholo_connection} of Claim \ref{claim1} we can rewrite the constraints using angular momentum $\Omega$ as 
\begin{align}
\begin{bmatrix}
\xi_{1} \\
\xi_{2} \\
\xi_{3}\\
\xi_{4}
\end{bmatrix} = 
\begin{bmatrix}
-A_{1} \dot{\alpha} + \Omega_{1} \\
\Omega_{2} \\
\Omega_{3} \\ 
-A_{2} \dot{\alpha} + \Omega_{4}
\end{bmatrix}
\end{align}
with $A_{1} = (r m_{b}b \cos \alpha )/h$ and $A_{2} = (m_{b}b \cos \alpha)/h$. Using Theorem \ref{thm2}, the reduced Lagrangian in terms of $(\alpha,\dot{\alpha}, \Omega)$ 
\begin{align*}
l &(\alpha,  \dot{\alpha},\Omega) = \frac{1}{2}(m_{b} + 2m_{W})[\Omega_{1}^{2} + A_{1}^{2} \dot{\alpha}^{2} - 2A_{1} \Omega_{1}\dot{\alpha} +\Omega^{2}_{2}] + \frac{1}{2} f(\alpha) \Omega^{2}_{3} \\
& + \frac{1}{2}(m_{b}b^{2} + I_{B}) \dot{\alpha}^{2} + \frac{1}{2}I_{Wyy}2[\Omega^{2}_{4} + A_{2}^{2}\dot{\alpha}^{2} - 2A_{2}\Omega_{4}\dot{\alpha}] \\
& + m_{b}b\sin \alpha \Omega_{3} \Omega_{2} + m_{b}b \cos \alpha \dot{\alpha} \Omega_{1} - m_{b}b \cos \alpha A_{1} \dot{\alpha}^{2} - m_{b}gb \cos \alpha 
\end{align*}
The constraint space $\mathfrak{b}^{\mathfrak{S}}$ is defined by
\begin{equation}
\mathfrak{b}^{\mathfrak{S}} = \{ \Omega_{1} = r\Omega_{4} \quad \mbox{and} \quad \Omega_{2}=0 \}
\end{equation}
Substituting the above constraints, the constraint reduced Lagrangian is obtained as
\begin{align}\nonumber
\begin{split}
l_{c}(\alpha, \dot{\alpha}, \Omega)&  = \frac{1}{2}h \Omega_{4}^{2} + \frac{1}{2} f(\alpha) \Omega^{2}_{3} + \frac{1}{2}(m_{b}b^{2} + I_{B} - \frac{rm_{b}^{2}b^{2} \cos^{2} \alpha}{h}) \dot{\alpha}^{2}  - m_{b}gb \cos \alpha , \\
& = \frac{1}{2} \left( \frac{p_{1}^{2}}{h} + \frac{p_{2}^{2}}{f(\alpha)} \right) + \frac{1}{2}(m_{b}b^{2} + I_{B} - \frac{m_{b}^{2}b^{2} \cos^{2} \alpha}{h}) \dot{\alpha}^{2}  -m_{b}gb \cos \alpha .
\end{split}
\end{align}
The shape dynamics is calculated from \eqref{shape_dynamics}(also refer Ostrowski (1996) \cite{ostrowski_thesis}, \cite{zenkov}) as 
\begin{equation}\label{eqn:3}
\frac{d}{dt} \frac{\partial l_{c}}{\partial \dot{\alpha}} - \frac{\partial l_{c}}{\partial \alpha} = \langle\mbox{ad}^{\ast}_{\xi} \frac{\partial l}{\partial \xi}, A(\cdot) \rangle + \langle \frac{\partial l}{\partial \xi}, \left( \mbox{d}A(\dot{\alpha}, \cdot) + \frac{\partial \Gamma(\alpha)p}{\partial \alpha} \right) \rangle ,
\end{equation}
where
\begin{equation}\nonumber
\mbox{d}A(\dot{\alpha}, \cdot) = \frac{\partial A}{\partial \alpha}\dot{\alpha} - \frac{\partial A \dot{\alpha}}{\partial \alpha} = 0;
\end{equation}
\begin{equation}\nonumber
\mbox{ad}^{\ast}_{\xi}\frac{\partial l}{\partial \xi} = \begin{bmatrix}
 - \frac{\partial l}{\partial \xi_{2}} \xi_{3} \\ \frac{\partial l}{\partial \xi_{1}} \xi_{3} \\ \xi_{2} \frac{\partial l}{\partial \xi_{3}} - \xi_{1} \frac{\partial l}{\partial \xi_{3}} \\ 0
\end{bmatrix};
\frac{\partial l}{\partial \xi} = \begin{bmatrix}
(m_{b} + 2m_{W})\xi_{1} - m_{b}b \cos \alpha \dot{\alpha} \\
 - m_{b}b \sin \alpha \xi_{3} + (m_{b} + 2m_{W}) \xi_{2} \\
-m_{b}b \sin \alpha \xi_{2} + f(\alpha) \xi_{3} \\
I_{Wyy}\frac{2}{r^{2}} \xi_{4}
\end{bmatrix}
\end{equation}

Substituting these terms and calculating the partials we get the resultant shape dynamics as \eqref{shape_dynamics_eqn}. The momentum equations are calculated in preceding section and is given by equation \eqref{momentum_dyn1} and \eqref{momentum_dyn2}. $ \hfill \square$ \\
To summarize the idea presented so far for the WIP, we have used the $G_2$ symmetry group and the tools associated to simplify the dynamics to the form:
\begin{align}
& g^{-1}\dot{g} + \mathcal{A}(\alpha)\dot{\alpha} = \Gamma(\alpha)p, \label{eqn:5} \\
& \dot{p}_{1} = \frac{m_{b} r b \sin \alpha}{ [f(\alpha)]^{2}} p_{2}^{2} + u_{1} , \label{eqn:6}  \\
& \dot{p}_{2} = - \frac{m_{b} b r \sin \alpha p_{2}}{ f(\alpha) h } \left[ m_{b} b \cos \alpha \dot{\alpha} + p_{1} \right] + u_{2}, \label{eqn:7} \\
& \left( m_{b}b^{2} + I_{B} - \frac{m_{b}^{2}b^{2}\cos^{2} \alpha}{h} \right) \ddot{\alpha} = - \frac{m_{b} b \sin 2 \alpha}{2h} \dot{\alpha}^{2} \nonumber \\
& \qquad + \frac{1}{2} \frac{\left(m_{b}^{2}b^{2}r \sin 2 \alpha - h f^{\prime}(\alpha) \right)}{h f(\alpha)^{2}}p_{2}^{2}  - m_{b}gb \sin \alpha  \label{eqn:8}
\end{align}
Equation \eqref{eqn:6} is the momentum conjugate to the forward velocity and \eqref{eqn:7} is the yaw momentum. The controls $u_{1}$ and $u_{2}$ are the torque responsible for the forward motion and the yaw motion of the WIP. In particular, $u_{1} = \frac{r(\tau_{1} + \tau_{2})}{2}$ and $u_{2}=\frac{r(\tau_{2}-\tau_{1})}{d}$, where $\tau_{1}$ and $\tau_{2}$ are the wheel torques.

Thus we have reduced the equation from 3 second order equations with 3 constraints equations to 5 first order equations and 1 second order equation (7 first order equations). The process of reconstruction is done by lifting the shape curve through the Lie algebra through connection \eqref{eqn:5} in order to solve the dynamics on the fiber. Control laws could now be synthesized 
on this set of equations, with the objective of preserving a vertically upright position $(\alpha =0)$ and a constant 
forward momentum $p_1$ and yaw momentum $p_2$.
\section{Future work}
With the dynamics given by  \eqref{eqn:6}-\eqref{eqn:8}, the controlled directions 
are given by the fibers of a subbundle $\mathcal{F}^{\ast}$ of the momentum phase space $T^{\ast}Q$. This $\mathcal{F}^{\ast}$ is a subset of $\mathcal{S}^{\ast}$, where $\mathcal{S}^{\ast} \subset T^{\ast}Q $ is a bundle over $Q$ whose fibers are the dual of the fibers of $\mathcal{S}$. 
 Sometimes, this may fail to be a cotangent space to the submanifold of the configuration space, which is typical when control torques are used. In such a situation, a suitable choice of basis
for the Lie algebra may result in simpler controlled dynamics. In \cite{bloch2009quasivelocities}, the authors discuss quasivelocities and its application in control, used Hamel's equation and body frame basis to derived the momentum equations in the body frame and prove conservation laws. Moreover, such a change in frame (Lie algebra basis) may assist in design stabilizing inputs for the system. The future work is to examine such a change in basis and design control laws appropriately.
\section{Acknowledgements}
The authors thank the support of DST-India and DAAD-Germany for financial support of this work under the aegis of a collaborative exchange project between TU Munich and IIT-Bombay. The second author performed a part of this work while he was a Pratt and Whitney Visiting Professor in the
Department of Aerospace Engineering at the Indian Institute of Science, Bangalore, India.
\section{Appendix}
Symmetries plays an important role in nonholonomic systems. Suppose we are given a Lagrangian $L$ and a smooth constraint nonintegrable distribution $\mathcal{D}$, then the action of a group and symmetry are defined as follows:
\begin{defn}
The (left) \textit{action} of a Lie group $G$ on a smooth manifold $Q$ is a smooth mapping $\Phi : G\times Q \longrightarrow Q$, such that 1) $\Phi(e,q)= q$ for all $q \in M$ and $\Phi(g,\Phi(h,q))=(gh,q)$ for all $g,h \in G$ and $q\in M$, and 2) For every $g\in G$, the map $\Phi_{g}$ is diffeomorphism.
\end{defn}
\begin{defn}
The tangent lift of a group action $\Phi$ is
\begin{align}\nonumber
\begin{split}
& T\Phi : G \times TQ \longrightarrow TQ \\
& (g,(q,\dot{q}))\rightarrow T\Phi(g,q,\dot{q}) = (\Phi(g,q),T_{q}\Phi(g,\dot{q})).
\end{split}
\end{align}
\end{defn}
\begin{defn}
A function $F$ is invariant (or symmetric) with respect to an action $\Phi$ of a Lie group $G$ if, for every $g\in G$, the map $\Phi_{g}$ is a symmetry of $F$, that is, $F\circ \Phi_{g}=F$. The group $G$ is then called a symmetry group of $F$.
\end{defn}
\begin{defn}
A distribution $\mathcal{D}$ is invariant in a sense that the action of $g \in G$ maps $\mathcal{D}_{q}$ at point $q$ to $\mathcal{D}_{gq}$ at point $gq$.
\end{defn}
In the context here, a mechanical system is invariant under a Lie group action $G$ if the system Lagrangian $L$ and constraint distribution $\mathcal{D}$ are invariant. If this holds, then the system posses group symmetry.
\\
\textit{Orbit Space:} The action of a Lie group $G$ through a point on manifold forms a group orbit. The group orbit through a point $q$ on $Q$ is defined as
\begin{equation}
\mbox{Orb}(q) \triangleq \{gq ~| g \in G \}.
\end{equation}
Let $\mathfrak{g}$ be the Lie algebra of the Lie group $G$ and $\xi \in \mathfrak{g}$. 
\begin{defn}
Suppose $\Phi: G \times Q \longrightarrow Q$ is an action. For $\xi \in \mathfrak{g}$, the map $\Phi^{\xi}: \mathbb{R} \times Q \longrightarrow Q$ defined as $\Phi^{\xi}(t,q) = \Phi(exp(\xi t),q)$ is an $\mathbb{R}$ action (called as $flow$) on $Q$ . The vector field that generates this flow, is given by
\begin{equation}
 \xi_{Q}(q) = \frac{d}{dt} \Big|_{t=0} \Phi(\mbox{exp}(t\xi),q)
\end{equation}
and is called an infinitesimal generator of the action corresponding to $\xi$.
\end{defn}
Hence, the tangent space to a group orbit through a point $q$ is given by the set of infinitesimal generators at that point, denoted by $T_{q}\mbox{Orb}(q) = \{ \xi_{Q}(q) ~ | ~ \xi \in \mathfrak{g} \}$. 
%
%
%
Now, let us revisit our bundle structure.

Consider a free and proper action of a Lie group $G$ on $Q$. The quotient space (shape space) $ Q/G \cong S$ and the projection map $\pi : Q \longrightarrow Q/G$ define a bundle structure referred to as a principal bundle and the kernel of $T_{q}\pi$, which is the tangent space to the group orbit, is called as the vertical space of the bundle at point $q$. We now define a connection, termed as principal connection on $Q(S,G,\pi)$, as follows.
%
\begin{defn}
A principal connection on $Q(S,G,\pi)$ is a $\mathfrak{g}$ valued 1-form $\mathcal{A}$ on $Q$ satisfying, 
\begin{enumerate}
\item $\mathcal{A}(\xi_{Q}(q)) = \xi$, $\forall \xi \in \mathfrak{g}$ and $q \in Q$;
\item $\mathcal{A}(\Phi_{\ast}X_{p}) = \mbox{Ad}_{g}\mathcal{A}(X_{p})$ for all $X_{p} \in TQ$, where $\mbox{Ad}$ denotes the adjoint of $G$ on $\mathfrak{g}$. This is called the equivariance of the connection.
\end{enumerate}
\end{defn}
Then the horizontal subspace can be denoted as $H_{q} = \{ v_{q} | \mathcal{A}(v_{q}) =0  \} $, and $\mathcal{A}$ can be
expressed  using the above property as $\mathcal{A}(g,r,\dot{g},\dot{r}) = \mbox{Ad}_{g}(\xi + A(r)\dot{r})$. Here, $A$ is the \textit{local form of connection} $\mathcal{A}$ which only depends on $r$. \\
Assume that the system Lagrangian $L$ and the constraint distribution $\mathcal{D}$ of a constrained mechanical system are invariant under a Lie group action $G$. Then following \cite{ostrowski_thesis},\cite{cortes}, assume $\mathcal{D}_{q} + T_{q} (\mbox{Orb})(q) = TQ$. Since we are interested in those symmetry directions which are compatible with the constraints, let $\mathcal{S}_{q} = \mathcal{D}_{q} \cap T_{q}\mbox{Orb}(q)$ be the intersection of the tangent space to the orbit with the constraint distribution. $\mathcal{S}_{q}$ is called the constraint fiber distribution and the union of these spaces over $q \in Q$ forms a vector bundle $\mathcal{S}$ over $Q$. The Lie algebra for this intersection is defined as below:
\begin{defn}[Lie algebra of $\mathcal{S}_{q}$]
Define for each $q \in Q$ the subspace $\mathfrak{g}^{q}$ to be the set of Lie algebra elements in $\mathfrak{g}$ whose infinitesimal generators evaluated at $q$ lie in both $\mathcal{D}_{q}$ and $T_{q}\mbox{Orb}(q)$: $\mathfrak{g}^{q} = \{ \xi \in \mathfrak{g} \quad | \xi_{Q}(q) \in \mathcal{S}_{q} \}$. 
%
The corresponding bundle over $Q$ whose fiber at point $q$ is $\mathfrak{g}^{q}$ is denoted by $\mathfrak{g}^{\mathcal{D}}$.
\end{defn}
\subsection{Nonholonomic momentum and reduced dynamics}
For the bundle $\mathfrak{g}^{\mathcal{D}} \longrightarrow Q$ whose fiber at a point $q$ is given by $\mathfrak{g}^{q}$, the nonholonomic momentum is defined as
\begin{align}\label{momentum_formula}
& J^{nhc} : TQ \longrightarrow \mathfrak{g}^{\mathcal{D}^{\ast}} \hspace{1.5cm}  (q,\dot{q}) \mapsto J^{nhc}(q,\dot{q}) \nonumber  \\
& J^{nhc}(q,\dot{q})(\cdot) : \mathfrak{g}^{\mathcal{D}} \longrightarrow \mathbb{R}, \hspace{1cm} \xi^{q} \mapsto \langle \frac{\partial L}{\partial \dot{q}}, \xi^{q}_{Q}(q) \rangle.
\end{align}
%
The nonholonomic momentum provides the nonholonomic connection, and the horizontal space at point $q$ of this connection is the orthogonal complement of $S$ lying in the constraint distribution, i.e, $H_{q} = S_{q}^{\bot} \cap \mathcal{D}_{q}$. Further, consider $e_{1}(q), \cdots, e_{s}(q),e_{s+1}(q), \cdots e_{k}(q)$ to be the $q$-dependent Lie algebra basis of $\mathfrak{g}$ such that the first `$s$' element span the subspace $\mathfrak{g}^{q}$ . Let $e_{1}(r), \cdots, e_{s}(r),e_{s+1}(r), \cdots e_{k}(r)$ 
denote this basis of $\mathfrak{g}$ at $g=Id$. Then the body fixed basis are defined as $e_{k}(q) = Ad_{g}e_{k}(r)$. Thus, in this basis the momentum is defined as
\begin{equation}\label{momentum_formula1}
\langle J^{nhc}(g,r,\dot{g},\dot{r}), e_{i}(g,r) \rangle = \langle \frac{\partial l}{\partial \xi}, e_{i}(r) \rangle := p_{i} , \quad \quad 1 \leq i \leq s.
\end{equation} 
where $l$ is the reduced Lagrangian of the system and $p_{i}$ denotes the momentum in the body frame. 
\begin{defn}[Momentum equation \cite{BKMM}]
The generalized momentum equation in body representation on the principle bundle $Q \longrightarrow Q/G$ is given as
\begin{equation}\label{general_moment}
\frac{d}{dt} p_{i} = \langle \frac{\partial l}{\partial \xi}, [\xi, e_{i}(r)] + \frac{\partial e_{i}}{\partial r} \dot{r} \rangle .
\end{equation}
\end{defn}
Now, consider a map $A^{s} : T_{q}Q \longrightarrow S_{q}$, such that $(q,\dot{q}) \mapsto (\bar{I}^{-1}J^{nhc})_{Q}$,
%
where $\bar{I}: \mathfrak{g}^{\mathcal{D}} \longrightarrow \mathfrak{g}^{\mathcal{D}^{\ast}}$ is the local locked inertia tensor in the group direction. And $A^{kin}: T_{q}Q \longrightarrow S_{q}^{\bot}$ is the kinematic connection orthogonal to the kinetic energy metric. %
 Then the constraints and momentum equations are written as
\begin{equation}
A^{kin}(q)(\dot{q}) =0; \quad \quad A^{s}(q)(\dot{q}) = (\bar{I}^{-1}p)_{Q}.
\end{equation}
where $p = \langle J^{nhc}(q)\dot{q}, \xi^{q} \rangle $. Therefore the nonholonomic connection is the sum of kinematic and symmetric connection, $A^{nhc} = A^{kin} + A^{s}$. \\
\textbf{Reduced dynamics:} If $l$ the Lagrangian reduced by the group action and $p_{b}$ is the body momenta in the group directions in the constraint manifold, then the constraint can be written as, 
$ \bar{I}^{-1}p = \xi +  \mathcal{A}(r) \dot{r}$, 
where $\mathcal{A}$ denotes the local form of the nonholonomic connection. The constrained Lagrangian, expressed
in the momentum variable $p$ is,
\begin{equation}
l_{c}(r,\dot{r}, p) = l(r, \dot{r},\xi) \big |_{\xi = -\mathcal{A}(r) \dot{r} + \bar{I}^{-1}p}
\end{equation}
Expressing $\Omega \triangleq \bar{I}^{-1}p$ which is the body angular velocity in a body fixed basis at identity, we write the reduced constrained Lagrangian as
\begin{equation}
l_{c}(r,\dot{r}, \Omega) = l(r,\dot{r}, -\mathcal{A}(r)\dot{r} + \Omega)
\end{equation}
Then by the reduced nonholonomic constrained variational principle, the equations of motion are given by
\begin{eqnarray}\label{shape_dynamics}
\frac{d}{dt} \frac{\partial l_{c}}{\partial \dot{r}} - \frac{\partial l_{c}}{\partial r} = N(r,\dot{r},p),
\end{eqnarray}
where
\begin{equation}\nonumber
N = \langle\mbox{ad}^{\ast}_{\xi} \frac{\partial l}{\partial \xi}, \mathcal{A}(\cdot) \rangle + \langle \frac{\partial l}{\partial \xi}, \left( \mbox{d}A(\dot{r}, \cdot) \frac{\partial \Gamma(r)p}{\partial r} \right) \rangle,~ \mbox{with } \mbox{d}A = \frac{\partial \mathcal{A}(r)}{\partial r}\dot{\alpha} - \frac{\partial \mathcal{A}(r) \dot{r}}{\partial r}
\end{equation}
%
where $\Gamma : \mathfrak{g}^{\mathcal{D}^{\ast}} \longrightarrow \mathfrak{g}^{\mathcal{D}}$ is the inverse reduced inertia tensor. Hence, equation \eqref{general_moment} and \eqref{shape_dynamics}, together with the reconstruction equation $g^{-1}\dot{g} =  - \mathcal{A}(r) \dot{r} + \Gamma(r)p$, 
%
give the complete dynamics of the system termed as the reduced nonholonomic Lagrange-d'Alembert-Poincar\'{e} equation of motion \cite{BKMM}.



\end{document}